\documentclass[AMA,STIX1COL]{WileyNJD-v2}
\usepackage{moreverb}

\newcommand\BibTeX{{\rmfamily B\kern-.05em \textsc{i\kern-.025em b}\kern-.08em
T\kern-.1667em\lower.7ex\hbox{E}\kern-.125emX}}

\articletype{Research Article}%

\received{<day> <Month>, <year>}
\revised{<day> <Month>, <year>}
\accepted{<day> <Month>, <year>}

\usepackage{setspace}
\usepackage{color}
\usepackage{amsmath}
\usepackage{amsfonts}
\usepackage{verbatim}
\usepackage{amsmath}
\usepackage{graphicx}
\usepackage[caption=false]{subfig}%
\providecommand{\U}[1]{\protect\rule{.1in}{.1in}}

\newcommand{\be}{\begin{equation}}
\newcommand{\ee}{\end{equation}}

\newcommand{\mincir}{\raise
-3.truept\hbox{\rlap{\hbox{$\sim$}}\raise4.truept\hbox{$<$}\ }}
\newcommand{\magcir}{\raise
-3.truept\hbox{\rlap{\hbox{$\sim$}}\raise4.truept\hbox{$>$}\ }}

\RequirePackage{latexsym}
\usepackage{hyperref}
\hypersetup{
   colorlinks=true,
    linkcolor=red,
   filecolor=magenta,      
    urlcolor=cyan,
    pdfpagemode=FullScreen}

\begin{document}

\title{$f(T,B)$ gravity in a Friedmann--Lema\^{\i}tre--Robertson--Walker universe with nonzero spatial
curvature}

\author[1,2]{Andronikos Paliathanasis}

\author[3,1]{Genly Leon*}

\authormark{ANDRONIKOS PALIATHANASIS \& GENLY LEON	}

\address[1]{Institute of Systems Science, Durban University of Technology, PO Box 1334,
Durban 4000, South Africa}

\address[2]{Instituto de Ciencias F\'{\i}sicas y Matem\'{a}ticas, Universidad Austral de
Chile, Valdivia 5090000, Chile}

\address[3]{Departamento de Matem\'aticas, Universidad Cat\'olica del Norte, Avda. Angamos 0610, Casilla 1280 Antofagasta, Chile}

\corres{*Genly Leon, Departamento de Matem\'aticas, Universidad Cat\'olica del Norte, Avda. Angamos 0610, Casilla 1280 Antofagasta, Chile \& Institute of Systems Science, Durban University of Technology, PO Box 1334, Durban 4000, South Africa \email{genly.leon@ucn.cl}}

\abstract[Abstract]{We investigate exact solutions and the asymptotic dynamics for the Friedmann--Lema\^{\i}tre--Robertson--Walker universe with nonzero spatial curvature in the fourth-order modified teleparallel gravitational theory known as $f\left(  T, B\right)  $ theory. We show that the field equations admit a minisuperspace description, and they can reproduce any exact form of the scale factor. Moreover, we calculate the equilibrium points and analyze their stability. We show that Milne and Milne-like solutions are supported, and the de Sitter universe is provided. To complete our analysis, we use Poincaré variables to investigate the dynamics at infinity.}


\jnlcitation{\cname{%
\author{A. Paliathanasis}, and \author{G. Leon}} (\cyear{2022}), 
\ctitle{$f(T,B)$ gravity in a Friedmann--Lema\^{\i}tre--Robertson--Walker universe with nonzero spatial
curvature},   \cjournal{Math Meth Appl Sci.} 2022;1-18. \url{https://doi.org/10.1002/mma.8728}}
\date{\today}
\maketitle

\section{Introduction}

\label{sec1}

Cosmologists have widely studied alternative and modified theories of gravity in recent years \cite{clifton, sd, Ishak:2018his} because geometrodynamical
degrees of freedom are introduced into the gravitational Action Integral to explain recent cosmological observations
\cite{Teg, Kowal, Komatsu, Ade15,sn1,sn2}. In particular, geometric invariants are
used to modify the Einstein-Hilbert Action.  See for instance
\cite{clifton,sd,sd1,sd2,sd3} and references therein.

The fundamental geometric invariant function of General Relativity is the Ricciscalar $R$.  
The Levi-Civita connection defines $R$. However,
Einstein showed that if the fundamental connection curvature-less
Weitzenb{\"{o}}ck connection \cite{Weitzenb23} and the torsion scalar $T$ are
used for the definition of the gravitational theory, then the resulting theory
is equivalent to General Relativity known as the teleparallel equivalence of
General Relativity (TEGR) \cite{ein28,cc}. There is a plethora of modified
theories inspired by teleparallelism with many interesting results in
cosmology and astrophysics \cite{fer1,fer2,fer4,fer5,fer7,f01,mm1,mm3,mm4}. 
For reviews in teleparallelism, we refer the reader to references \cite{rew1,rew2}.

We are interested in the fourth-order teleparallel theory of gravity known as
the $f\left(  T, B\right)  $ theory. The gravitational Action Integral is a defined
by an arbitrary function $f$ of the torsion scalar $T$ and of the boundary
term $B$, which is related to the torsion scalar and the Ricciscalar, that is
$B=R+T$. The theory was introduced in detail in \cite{bahamonde}. However, a
similar fourth-order teleparallel theory was introduced before in
\cite{myr11}. There are various studies in the literature on $f\left(
T,B\right)  $ theory. The cosmological dynamics in a spatially flat
Friedmann--Lema\^{\i}tre--Robertson--Walker (FLRW) universe were investigated
in detail in a series of works \cite{an1,an1a,an1b,an1c} where it was found
that for a plethora of functions $f$ the modified teleparallel theory can
describe the main eras of the cosmological history. Exact and analytic
solutions were found in \cite{an2,an3}, while some bouncing solutions were
determined in \cite{ftb02}. Recently, a new inhomogeneous exact solution was
derived in \cite{ss1}. The quantization of $f\left(  T, B\right)  $ by
using the minisuperspace description to write the Wheeler-DeWitt equation of
quantum cosmology was studied in \cite{min1}. Cosmological constraints of
$f\left(  T,B\right)  $ theory can be found in \cite{ftb03,ftb02a,Escamilla-Rivera:2019ulu} while some
astrophysics applications are presented in \cite{ftb01,fbb,nr1,nr2,nr3,nr4}.
Anisotropic spacetimes in $f\left(  T, B\right)  $  theory investigated before in \cite{anan1,anan2,anan3}. 
Specifically, the dynamical evolution for the physical parameters investigated in the case of 
Bianchi I \cite{anan1}, Kantowski-Sachs \cite{anan2}, and Bianchi III \cite{anan3} background geometries. 

In the following, we consider in $f\left(  T, B\right)  $ theory in the background
space of an FLRW universe with nonzero spatial curvature. It has been
found that the inflationary scenario is not affected by the presence of
negative curvature in the background space \cite{in2,in3}. Thus, such an
analysis is important for studying teleparallelism in the very early
stages of the universe. We investigate the existence of power-law and
exponential scale factors in the $f\left(  T, B\right)  $ theory, which can describe
inflation. An accelerated expansion of the universe described by the exponential scale
factor solves various problems in cosmology, such as the \textquotedblleft
flatness\textquotedblright, \textquotedblleft horizon\textquotedblright\ and
monopole problems \cite{f1,f2}. A recent study on the effects of curvature in
teleparallelism was performed in \cite{ftc} where Milne and Milne-like universes
are supported in $f\left(  T\right)  $ theory. At the same time, the de Sitter expansion
is provided by the $f\left(  T\right)  $ theory without the necessity to introduce
a cosmological constant term. In \cite{ftcur} bouncing solutions in $f\left(
T\right)  $ cosmology with nonzero spatial curvature was studied.

Furthermore, we investigate the global dynamics of the field equations by
performing a complete dynamical analysis \cite{alc}. Indeed, we determine the
equilibrium points and investigate their stability properties. At every
equilibrium point, the scale factor is described by an exact solution 
corresponding to a specific epoch of the cosmological history \cite{dn1,dn2,dn3}.
The analysis of the asymptotic behaviour for the theory is essential for a
better understanding of the viability of the model \cite{lc1,lc2}.

The plan of the paper is as follows.

In Section \ref{sec2} we present the basic properties of teleparallelism and
we give the gravitational field equations for the $f\left(  T,B\right)  $
theory. Moreover, for the case of FLRW geometries, we derive the minisuperspace
description for the field equations and write the point-like Lagrangian
using a Lagrange multiplier. For the case of $f\left(  T,B\right)
=T+F\left(  B\right)  $ in Section \ref{sec3} we prove that the gravitational
theory supports exact solutions of interest, specifically those with power-law and exponential scale factor. The stability
properties of these important solutions are investigated in Section
\ref{sec4}. In particular, we investigate the global evolution of the field
equations by investigating the equilibrium points and their stability properties.
We discuss our results in\ Section \ref{sec5}.

\section{$f\left(  T,B\right)  $ gravity}

\label{sec2}

The fundamental geometric objects in teleparallelism are the vierbein fields
${\mathbf{e}_{\mu}(x^{\sigma})}$. The vierbein fields introduce the dynamical
variables of the theory, and they form an orthonormal basis for the tangent
space at each point $P$ such that $g(e_{\mu},e_{\nu})=\mathbf{e}_{\mu}%
\cdot\mathbf{e}_{\nu}=\eta_{\mu\nu}$, where $\eta_{\mu\nu}~$is the line
element of the Minkowski spacetime, $\eta_{\mu\nu}=\text{diag}\left(  -,+,+,+\right)
$.

Furthermore, for the vierbein fields, it holds that
\begin{equation}
\ [e_{\mu},e_{\nu}]=c_{\nu\mu}^{\beta}e_{\beta}~\text{\ where }c_{\left(
\nu\mu\right)  }^{\beta}=0. \label{cc.01}%
\end{equation}

In general, in the nonholonomic coordinates, the covariant derivative
$\nabla_{\mu}$ is defined by the nonsymmetric connection
\begin{equation}
\mathring{\Gamma}_{\nu\beta}^{\mu}=\{_{\nu\beta}^{\mu}\}+\frac{1}{2}%
g^{\mu\sigma}(c_{\nu\sigma,\beta}+c_{\sigma\beta,\nu}-c_{\mu\beta,\sigma}),
\label{cc.02}%
\end{equation}
in which $\{_{\nu\beta}^{\mu}\}$ is the symmetric Levi-Civita connection of
Riemannian geometry.

When $\mathbf{e}_{\mu}\cdot\mathbf{e}_{\nu}=\eta_{\mu\nu}$,~it follows that
\cite{Weitzenb23}%
\begin{equation}
\mathring{\Gamma}_{\nu\beta}^{\mu}=\frac{1}{2}\eta^{\mu\sigma}(c_{\nu
\sigma,\beta}+c_{\sigma\beta,\nu}-c_{\mu\beta,\sigma}), \label{cc.03}%
\end{equation}
where now $\mathring{\Gamma}_{\nu\beta}^{\mu}$ describes the Ricci rotation
coefficients. $\mathring{\Gamma}_{\nu\beta}^{\mu}$ has the property that it is
antisymmetric in the first two indices, i.e. $\mathring{\Gamma}_{\mu\nu\beta
}=-\mathring{\Gamma}_{\nu\mu\beta}$, with $\mathring{\Gamma}_{\mu\nu\beta
}=\eta_{\mu\sigma}\mathring{\Gamma}_{\nu\beta}^{\mu}$.

Consequently, the non-null torsion tensor can be defined as $T_{\mu\nu}^{\beta
}=\mathring{\Gamma}_{\nu\mu}^{\beta}-\mathring{\Gamma}_{\mu\nu}^{\beta},$ with
the scalar $\ T={S_{\beta}}^{\mu\nu}{T^{\beta}}_{\mu\nu}$. The geometric
object ${S_{\beta}}^{\mu\nu}$ is given by the expression
\begin{equation}
{S_{\beta}}^{\mu\nu}=\frac{1}{2}({K^{\mu\nu}}_{\beta}+\delta_{\beta}^{\mu
}{T^{\theta\nu}}_{\theta}-\delta_{\beta}^{\nu}{T^{\theta\mu}}_{\theta}),
\label{cc.04}%
\end{equation}
with $K_{~~~\beta}^{\mu\nu}=-\frac{1}{2}({T^{\mu\nu}}_{\beta}-{T^{\nu\mu}%
}_{\beta}-{T_{\beta}}^{\mu\nu})$.

In the teleparallel equivalent of General Relativity, the fundamental scalar is
the torsion scalar $T$, instead of the Ricciscalar and the dynamical
variables are the vierbein fields, instead of the metric tensor, that is, the
gravitational Action Integral is
\begin{equation}
S_{T}=\frac{1}{16\pi G}\int d^{4}xeT, ~e=\det(e_{\mu}). \label{cc.05}%
\end{equation}

In this study, we are interested in an extension of the teleparallel theory known
as $f\left(  T,B\right)  $ gravity. Specifically, we consider the modified
Action Integral \cite{bahamonde,myr11}%
\begin{equation}
S_{f\left(  T,B\right)  }=\frac{1}{16\pi G}\int d^{4}xef\left(  T,B\right),
\label{cc.06}%
\end{equation}
where $B=2e^{-1}\partial_{\nu}\left(  eT_{\rho}^{~\rho\nu}\right)  $ corresponds to
the boundary term which relates the torsion scalar $T$ with the Ricciscalar
$R$, that is $R=-T+B~$\cite{bahamonde}. Function $f\left(  T,B\right)  $ is an arbitrary function which should be defined. In the case for which $f\left(
T,B\right)  $ is a linear function. Then, the field equations of the General Relativity are recovered, with or without the cosmological constant term.
Moreover, for $f\left(  T,B\right)  =f\left(  -T+B\right)  $, another
well-known theory of gravity is recovered, the so-called fourth-order modified
$f\left(  R\right)  $ theory.

The gravitational field equations follow from the variation of the Action
Integral (\ref{cc.06}) with respect to the vierbein fields. They are
\cite{bahamonde}%
\begin{align}
4\pi Ge\mathcal{T}_{a}^{\left(  m\right)  }{}^{\lambda}  &  =\frac{1}{2}%
eh_{a}^{\lambda}\left(  f_{,B}\right)  ^{;\mu\nu}g_{\mu\nu}-\frac{1}{2}%
eh_{a}^{\sigma}\left(  f_{,B}\right)  _{;\sigma}^{~~~;\lambda}+\frac{1}%
{4}e\left(  Bf_{,B}-f\right)  h_{a}^{\lambda}\,+(eS_{a}{}^{\mu\lambda})_{,\mu
}f_{,T}\nonumber\\
&  ~\ ~+e\left(  (f_{,B})_{,\mu}+(f_{,T})_{,\mu}\right)  S_{a}{}^{\mu\lambda
}~-ef_{,T}T^{\sigma}{}_{\mu a}S_{\sigma}{}^{\lambda\mu}, \label{cc.07}%
\end{align}
or equivalently%
\begin{align}
4\pi Ge\mathcal{T}_{a}^{\left(  m\right)  }{}^{\lambda}  &  =ef_{,T}%
G_{a}^{\lambda}+\left[  \frac{1}{4}\left(  Tf_{,T}-f\right)  eh_{a}^{\lambda
}+e(f_{,T})_{,\mu}S_{a}{}^{\mu\lambda}\right]  \nonumber \\
&  +\left[  e(f_{,B})_{,\mu}S_{a}{}^{\mu\lambda}-\frac{1}{2}e\left(
h_{a}^{\sigma}\left(  f_{,B}\right)  _{;\sigma}^{~~~;\lambda}-h_{a}^{\lambda
}\left(  f_{,B}\right)  ^{;\mu\nu}g_{\mu\nu}\right)  +\frac{1}{4}%
eBh_{a}^{\lambda}f_{,B}\right]  . \label{cc.08}%
\end{align}

\subsection{Minisuperspace description}

For the FLRW universe with nonzero curvature, that is, for the line element
\begin{equation}
ds^{2}=N^{ 2}dt^{2}-a^{2}\left(  t\right)  \left(  dr^{2}+\sin^{2}\left(
r\right)  \left(  d\theta^{2}+\sin^{2}(\theta) d\phi^{2}\right)  \right)
,~K=1,
\end{equation}
or%
\begin{equation}
ds^{2}=N^{ 2}dt^{2}-a^{2}\left(  t\right)  \left(  dr^{2}+\sinh^{2}\left(
r\right)  \left(  d\theta^{2}+\sin^{2}(\theta) d\phi^{2}\right)  \right)
,~K=-1,
\end{equation}
we consider the vierbein fields
\[
e_{\mu}=\left(N dt,a\left(  t\right)  E^{r}\left(  K\right)  ,a\left(
t\right)  E^{\theta}\left(  K\right)  ,a\left(  t\right)  E^{\phi}\left(
K\right)  \right),
\]
where the components of $\mathbf{E}\left(  K\right)  $ depend upon the spatial
curvature $K$ and, for simplicity, in the following, we set the lapse function $N=1$.

For the \textquotedblleft closed\textquotedblright\ universe, i.e. $K=1$, we
have \cite{ff1}
\begin{equation}
E^{r}\left(  K=1\right)  =-\cos(\theta) dr+\sin (r)\sin(\theta)\left(  \cos
(r) d\theta-\sin (r)\sin(\theta) d\phi\right),  \label{fc.03}%
\end{equation}%
\begin{align}
E^{\theta}\left(  K=1\right)   &  =\sin(\theta)\cos(\phi) dr-\sin(r)\left(\sin(r)\sin(\phi)-\cos(r)\cos(\theta)\cos(\phi)\right)  d\theta\nonumber \\
&  -\sin(r)\sin(\theta)\left(  \cos(r)\sin(\phi)+\sin(r)\cos(\theta)\cos(\phi)\right)
d\phi,  \label{fc.04}%
\end{align}%
\begin{align}
E^{\phi}\left(  K=1\right)   &  =-\sin(\theta)\sin(\phi) dr-\sin(r)\left(  \sin(
r)\cos(\phi)+\cos(r)\cos(\theta)\sin(\phi)\right)  d\theta\nonumber \\
&  -\sin(r)\sin(\theta)\left(  \cos(r)\cos(\phi)-\sin(r)\cos(\theta)\sin(\phi)\right)
d\phi. \label{fc.05}%
\end{align}
For the \textquotedblleft open\textquotedblright\ universe, i.e. $K=-1$,
it holds that \cite{ff1}%
\begin{equation}
E^{r}\left(  K=-1\right)  =\cos(\theta) dr+\sinh(r)\sin(\theta)\left(  -\cosh(r) d\theta+i\sinh(r)\sin(\theta) d\phi\right),  \label{fc.06}%
\end{equation}%
\begin{align}
E^{\theta}\left(  K=-1\right)   &  =-\sin(\theta)\cos(\phi) dr+\sinh(r)\left(
i\sinh(r)\sin(\phi)-\cos(r)\cos(\theta)\cos(\phi)\right)  d\theta\nonumber \\
&  +\sinh(r)\sin(\phi)\left(  \cosh(r)\sin(\phi)+i\sinh(r)\cos(\theta)\cos(\phi)\right)
d\phi,  \label{fc.07}%
\end{align}%
\begin{align}
E^{\phi}\left(  K=-1\right)   &  =\sin(\theta)\sin(\phi) dr+\sinh(r)\left(  i\sinh(r)\cos(\phi)+\cosh(r)\cos(\theta)\sin(\phi)\right)  d\theta\nonumber \\
&  +\sinh(r)\sin(\theta)\left(  \cosh(r)\cos(\phi)-\sinh(r)\cos(\theta)\sin(\phi)\right)
d\phi.  \label{fc.08}%
\end{align}

Thus, for this frame, the torsion scalar $T$~is calculated as \cite{ff1}%
\begin{equation}
T=6\left(  \frac{K}{a^{2}}-H^{2}\right),  \label{fc.10}%
\end{equation}
where$~H=\frac{\dot{a}}{a}~$is the Hubble function$~\dot{a}=\frac{da}{dt}$,
while the boundary term is%
\begin{equation}
B=-6\left(  \dot{H}+3H^{2}\right)  . \label{fc.11}%
\end{equation}

We introduce the Lagrange multipliers $\lambda_{1},~\lambda_{2}$. Thus the
Action Integral (\ref{cc.06}) for a the FLRW spacetime is written in the
equivalent form%
\begin{equation}
S_{f\left(  T,B\right)  }=\frac{1}{16\pi G}\int dt\left(  a^{3}f\left(
T,B\right)  -\lambda_{1}a^{3}\left(  T-6\left(  \frac{K}{a^{2}}-H^{2}\right)
\right)  -\lambda_{2}a^{3}\left(  B+6\left(  \dot{H}+3H^{2}\right)  \right)
\right)  . \label{fc.12}%
\end{equation}

Variation with respect to the variables $T$ and $B$ of (\ref{fc.12})
constrain the Lagrange multiplier. Indeed, from the equations of motion
$\frac{\delta}{\delta T}\left(  S_{f\left(  T,B\right)  }\right)  =0$ and
$\frac{\delta}{\delta B}\left(  S_{f\left(  T,B\right)  }\right)  =0$, it
follows that $\lambda_{1}=f_{,T}$ and $\lambda_{2}=f_{,B}$. Consequently,
expression (\ref{fc.12}) becomes%
\begin{equation}
S_{f\left(  T,B\right)  }=\frac{1}{16\pi G}\int dt\left(  a^{3}f\left(
T,B\right)  -f_{,T}\left(  a^{3}T-6\left(  aK-a\dot{a}^{2}\right)  \right)
-f_{,B}\left(  B+12a\dot{a}^{2}\right)  -6f_{,B}a^{2}\ddot{a}\right)  .
\label{fc.14}%
\end{equation}
Integration by parts of the last term of (\ref{fc.14}) gives%

\begin{equation}
\int dt\left(  6f_{B}a^{2}\ddot{a}\right)  =-\int dt\left(  12f_{,B}a\dot
{a}^{2}+6a^{2}f_{,BB}\dot{a}\dot{B}\right)  . \label{fc.15}%
\end{equation}

Thus, we can write the point-like Lagrangian function%
\begin{equation}
\mathcal{L}\left(  a,\dot{a},T,B,\dot{B}\right)  =-6f_{,T}a\dot{a}^{2}%
-6a^{2}f_{,BB}\dot{a}\dot{B}+a^{3}\left(  f-Tf_{,T}-Bf_{,B}\right)
+6aKf_{,T}, \label{fc.16}%
\end{equation}
which generates the gravitational field equations. We remark that
$\mathcal{L}\left(  a,\dot{a},T,B,\dot{B}\right)  $ provides field equations
of second-order.  However, $B$ has been introduced by a Lagrange multiplier and
includes the higher-order derivatives, such that $f\left(  T,B\right)$ theory is to be of fourth-order.

\subsection{$f\left(  T,B\right)  =T+F\left(  B\right)  $ theory}

{We proceed with our analysis by assuming the functional form $f\left(  T,B\right)
$ to be linear in $T,$ that is, $f\left(  T,B\right)  =T+F\left(  B\right)
$. A such function has been considered before in \cite{an1}. The main
mathematical novelty of this approach is that the point-like Lagrangian
(\ref{fc.16}) is regular, while for small values of function $F\left(
B\right)  $ we are very close to the limit of General Relativity. In such consideration, we assume a modification of the Action Integral for the TEGR, which follows from the existence of the boundary function $B$.}

{Now, we introduce  the new field, and potential definition  \begin{equation}
    \phi= F_{,B},  \; \text{and}\; \quad 
    V(\phi)= \left(F-B F_{,B}\right)/6. \label{Clairaut}
\end{equation}
Given an explicit form $V(\phi)$  we reconstruct $f(B)$ from the singular solution of Clairaut's equation \cite{Clairaut}.}

 {Therefore, the Lagrangian of the
field equations is%
\begin{equation}
\mathcal{L}\left(  a,\dot{a}, \phi,\dot{\phi}\right)  = \frac{1}{N}\left[-6 a\dot{a}^{2}%
-6a^{2} \dot{a}\dot{\phi} \right]+ N \left[6 a^{3} V\left(\phi\right)
+6aK\right], \label{fc.17} 
\end{equation}
whereby convenience, we have reinserted the lapse function $N$. }

Taking the variation of (\ref{fc.17}) with respect to $\{a, \phi, N\}$ we derive the field equations%
\begin{equation}
2\dot{H}+3H^{2}+\ddot{\phi}+3V\left(  \phi\right)  +Ka^{-2}=0,  \label{fc.18}%
\end{equation}%
\begin{equation}
\dot{H}+3H^{2}+V^{\prime}\left(  \phi\right)  =0,  \label{fc.19}%
\end{equation}
with constraint equation%
\begin{equation}
H^{2}+H\dot{\phi}+V\left(  \phi\right)  +Ka^{-2}=0,\label{fc.20}%
\end{equation}
 {where the equation \eqref{fc.18}  is provided by the Euler-Lagrange 
equations
with respect to the scale factor $a$,~$\frac{d}{dt}\frac{\partial\mathcal{L}%
}{\partial\dot{a}}-\frac{\partial\mathcal{L}}{\partial a}=0$. The 
scalar field equation \eqref{fc.19} arises from   
$\frac{d}{dt}\frac{\partial\mathcal{L}}{\partial\dot{\phi}%
}-\frac{\partial\mathcal{L}}{\partial\phi}=0$. Finally,  the equation  $\partial \mathcal{L}/\partial N|_{N=1}=0$ gives the Friedmann constrain \eqref{fc.20}. As usual, in all the above 
equations,  one can set $N=1$ after the derivations.} 

The field equations (\ref{fc.18}) and (\ref{fc.20}) can be written in the
equivalent form%
\begin{equation}
3H^{2}+ 3Ka^{-2}=\rho_{\phi},%
\end{equation}%
\begin{equation}
2\dot{H}+3H^{2}+Ka^{-2}=-p_{\phi},%
\end{equation}
where $\rho_{\phi}$ and $p_{\phi}$ are the cosmological fluid components which
correspond to the geometrodynamical degrees of freedom given by the nonlinear
$F\left(  B\right)  $ function. They are%
\begin{equation}
\rho_{\phi}=-3\left(  H\dot{\phi}+V\left(  \phi\right)  \right),
\end{equation}%
\begin{equation}
p_{\phi}=\ddot{\phi}+3V\left(  \phi\right).
\end{equation}
Thus, the equation of state parameter for the geometric fluid source is
defined as
\begin{equation}
w_{\phi}=-\frac{\frac{1}{3}\ddot{\phi}+V\left(  \phi\right)  }{H\dot{\phi
}+V\left(  \phi\right)  }.
\end{equation}
We observe that, when $\dot{\phi}\approx0$ and $\ddot{\phi}\approx0$, it
follows $w_{\phi}\simeq -1,$ that is, the limit of the cosmological constant is recovered.  {The dust matter domination is provided by $\ddot{\phi}+3V\left(  \phi\right) \approx 0$. When $V\left(  \phi\right) \simeq 0$, we have $w_{\phi}\simeq - \frac{1}{3}\ddot{\phi}  /(H\dot{\phi } )$. Therefore, radiation-dominated epoch corresponds to $V\left(  \phi\right) \simeq 0$ together with $ \ddot{\phi}  /(H\dot{\phi } ) \simeq -1$. That is, $V\left(  \phi\right) \simeq 0$  and $ d\ln(\dot{\phi } )/ d\ln a  \simeq -1$, i.e., $\dot{\phi}\simeq a^{-1}$, $ \phi \simeq \int a^{-1} dt$ mimics a radiation-dominated universe. These models of $f(T, B)$ gravity offer a unified description of the universe evolution (i.e. the matter era and the late-time acceleration epoch), similarly to the analysis of  \cite{Leon:2022oyy} in scalar-torsion theory. This analysis has been presented before in \cite{an1}, where this theory can describe various eras of cosmological history. However, the present work focuses on the existence of spatial curvature and how $f(T, B)$ gravity solves the flatness problem. }

In the following sections, we investigate the existence of exact solutions. Also, we study the cosmological dynamics for the field equations (\ref{fc.18}%
)-(\ref{fc.20}). Such an analysis provides important information that will help us understand spatial curvature's effects on the application of
teleparallelism in cosmology.

\section{Exact solutions}

\label{sec3}

We proceed by investigating the existence of exact solutions in which the scale
factor is a power-law function, i.e. $a\left(  t\right)  =a_{0}t^{p}$, the
exponential function, that is, $a\left(  t\right)  =a_{0}e^{H_{0}t}$, and the
Einstein-static universe, $a\left(  t\right)  =a_{0}$.

\subsection{Scaling solution}

For the scaling solution $a\left(  t\right)  =a_{0}t^{p}$, with $H\left(
t\right)  =pt^{-1}$, from the field equations (\ref{fc.18})-(\ref{fc.20}) we
find the linear second-order ordinary differential equation%
\begin{equation}
\ddot{\phi}-3pt^{-1}\dot{\phi}+2\left(  pt^{-2}+Kt^{-2p}\right)  =0
\end{equation}
with analytic solution%
\begin{equation}
\phi\left(  t\right)  =\phi_{0}+\frac{\phi_{1}}{1+3p}t^{1+3p}-\frac{2p}%
{1+3p}+\frac{K}{\left(  1-p\right)  \left(  1-5p\right)  }t^{2\left(
1-p\right)  }, ~p\notin \left\{1,\frac{1}{5},-\frac{1}{3}\right\},
\end{equation}%
\begin{equation}
\phi\left(  t\right)  =\phi_{0}+\frac{\phi_{1}}{4}t^{4}-\frac{1+K}{2}\ln
t, ~p=1,
\end{equation}%
\begin{equation}
\phi\left(  t\right)  =\phi_{0}+\frac{5}{8}\left(  \phi_{1}-\frac{5}%
{4}K\right)  t^{\frac{8}{5}}-\frac{1}{4}\left(  1-5Kt^{\frac{8}{5}}\right)
\ln t, ~p=\frac{1}{5},
\end{equation}%
\begin{equation}
\phi\left(  t\right)  =\phi_{1}+\phi_{1}\ln t-\frac{1}{3}\left(  \ln t\right)
^{2}+\frac{9}{32}Kt^{\frac{8}{3}}, ~p=-\frac{1}{3}.
\end{equation}

Similarly, for the scalar field potential $V\left(  \phi\left(  t\right)
\right)  $ we derive%
\begin{equation}
V\left(  \phi\left(  t\right)  \right)  =-\phi_{1}pt^{-1+3p}+\frac
{p^{2}\left(  1-p\right)  }{1+3p}t^{-2}-\frac{\left(  1-3p\right)  K}{\left(
1-5p\right)  }t^{-2p}~, ~p\notin \left\{1,\frac{1}{5},-\frac{1}{3}\right\},
\end{equation}%
\begin{equation}
V\left(  \phi\left(  t\right)  \right)  =-\phi_{1}t^{2}-\frac{K+1}{2}%
t^{-2}, ~p=1,
\end{equation}%
\begin{equation}
V\left(  \phi\left(  t\right)  \right)  =-\frac{\left(  \phi_{1}+5K\right)
}{5}t^{-\frac{2}{5}}+\frac{1}{100}t^{-2}-\frac{2}{5}t^{-\frac{2}{5}}\ln
t, ~p=\frac{1}{5},
\end{equation}%
\begin{equation}
V\left(  \phi\left(  t\right)  \right)  =\frac{3\phi_{1}-1}{9}t^{-2}-\frac
{3}{4}Kt^{\frac{1}{3}}-\frac{2}{9}t^{-2}\ln t, ~p=-\frac{1}{3}.
\end{equation}

Let us focus now on the case where $p=1$. The scale factor $a\left(  t\right)
=a_{0}t$, describes\ Milne (for $K=-1$) and Milne-like (for $K=1$)~universes.
For $K=-1$, it follows $\phi\left(  t\right)  =\phi_{0}+\frac{\phi_{1}}%
{4}t^{4}$ and $V\left(  \phi\right)  =-\phi_{1}t^{2}$, for which we observe that
$\rho_{\phi}=0$. Thus, there is not any contribution to the cosmological fluid
from the $F\left(  B\right)  $ component. On the other hand, for the $K=1$, and
the Milne-like solution, we observe that $\rho_{\phi}\neq0$. For large values
of $t$, $\phi\left(  t\right)  \simeq\phi_{0}+\frac{\phi_{1}}{4}t^{4}$ and
$V\left(  \phi\left(  t\right)  \right)  \simeq-\phi_{1}t^{2}-t^{-2}$, that
is,
\begin{equation}
V\left(  \phi\right)  \simeq-\left(  4\phi_{1}\left(  \phi-\phi_{0}\right)
\right)  ^{\frac{1}{2}}-\left(  \frac{4}{\phi_{1}}\left(  \phi-\phi
_{0}\right)  \right)  ^{-\frac{1}{2}}.
\end{equation}
Furthermore, for small values of $t$, that is, near to the initial
singularity, it follows $\phi\left(  t\right)  \simeq-\ln t$,~$V\left(
\phi\left(  t\right)  \right)  =-t^{-2}$, i.e.%
\begin{equation}
V\left(  \phi\right)  \simeq-e^{2\phi},
\end{equation}
 {which leads to 
\begin{align}   F \simeq \frac{1}{2} B \left(\ln
   \left(\frac{B}{12}\right)-1\right).
   \end{align}}
\subsection{Exponential scale factor solution}

Assume now the exponential scale factor $a\left(  t\right)  =a_{0}e^{H_{0}t}$.
Then the scalar field satisfies the second-order
ordinary differential equation,%
\begin{equation}
\ddot{\phi}-3H_{0}\dot{\phi}-2Ke^{-2H_{0}t}=0,
\end{equation}
with analytic solution%
\begin{equation}
\phi\left(  t\right)  =\phi_{0}+\frac{\phi_{1}}{3H_{0}}e^{3H_{0}t}+\frac
{K}{5H_{0}^{2}}e^{-2H_{0}t}.
\end{equation}

 {
For the potential function, we calculate%
\[
V\left(  \phi\right)  = 3H_{0}^{2}\left(-\frac{1}{3}+\frac{K}{5H_{0}^{2}%
}e^{-2H_{0}t}+\frac{\phi_{1}}{3H_{0}}e^{3H_{0}t}\right)=  3H_{0}^{2}\left(-\frac{1}{3}+ \phi - \phi_0\right).
\]
Thus, the scalar field
potential is described by the linear function $V\left(
\phi\right) = -\alpha + \beta \phi,$ $\alpha= \beta \left( \frac{1}{3}+ \phi_0\right)$, $\beta= 3H_{0}^{2}$ which leads to 
\begin{align}
F(B)= \left( \frac{1}{3}+ \phi_0\right)B. 
   \end{align} }
\subsection{Einstein-static universe}

For a static universe, $a\left(  t\right)  =a_{0}$, we calculate
\begin{align}
\phi\left(  t\right)   &  =Kt^{2}+\phi_{0}+\phi_{1}t, \\
V\left(  \phi\right)   &  =-K,
\end{align}
 {so  $F(B)=F_{1}B-6 K$, where $F_{1}$ is the integration constant.}

Below we continue our analysis by investigating the stability properties of the
above solutions and the asymptotic behaviour of the field equations.
Specifically, we study the dynamics of the field equations by determining the
equilibrium points and their stability.
\section{Dynamical analysis}

\label{sec4}

To study the dynamics of the field equations (\ref{fc.18}%
)-(\ref{fc.20}) we define the new variables \cite{kkd1}%
\begin{equation}	
x=\frac{\dot{\phi}}{\sqrt{H^{2}+\left\vert K\right\vert a^{-2}}}%
, ~y=\frac{V\left(  \phi\right)  }{H^{2}+\left\vert K\right\vert a^{-2}%
}, ~\eta=\frac{H}{\sqrt{H^{2}+\left\vert K\right\vert a^{-2}}}, ~\lambda
=\frac{V_{,\phi}\left(  \phi\right)  }{V\left(  \phi\right)  },
\end{equation}
 {which satisfies 
\begin{equation}
\left(1-  \text{sgn}(K)\right) \eta^2 + \text{sgn}(K)+\eta  x +y=0.
\end{equation}
Furthermore, we consider the new independent variable, $d\tau=\sqrt
{H^{2}+\left\vert K\right\vert a^{-2}}dt$, leading to the dynamical system 
\begin{align}
\frac{dx}{d\tau}& = \text{sgn}(K) \left(\eta^2-1\right)+\eta \left(3
   \eta +2 \eta^2 x +x \right)+y  (2 \lambda +\lambda
    \eta  x -3),\\
\frac{d y}{d\tau}& =y \left(\lambda  x+2 \eta  \left(2 \eta^2+\lambda y+1\right)\right), \\
\frac{d\eta}{d\tau}& = \left(\eta ^2-1\right) \left(2 \eta^2+\lambda  y\right),\\
  \frac{d \lambda}{d\tau} & = h x, 
\end{align}
where 
\begin{equation}
  h=  \frac{V_{,\phi \phi}}{V}-\frac{V_{,\phi}^2}{V^2}.  
\end{equation}}
In the new variables the
equation of state parameter for the effective fluid, $w_{tot}=-1-\frac{2}%
{3}\frac{\dot{H}}{H^{2}}$.  The deceleration parameter, $q=-1- \frac{\dot{H}}{H^{2}}$, is%
\begin{equation}
w_{tot}\left(  x,y,\eta,\lambda\right)  =1+\frac{2\lambda}{3\eta^{2}}y
\end{equation}
and%
\begin{equation}
q\left(  x,y,\eta,\lambda\right)  = 2 +\frac{\lambda  y}{\eta^2}.
\end{equation}

 {It is useful to define the quantity $\Omega_K=  \frac{|K|}{a^2 H^2}$, which is a dimensionless measure of the spatial curvature in an FLRW universe. Hence, although the analysis is for $K=1$ and $K=-1$ respectively, some points correspond to “flat FRW”, which is the situation where asymptotically $\Omega_K \rightarrow 0$. In other words, with the term flat, we mean asymptotically flat. Notice that $\eta^2=1/(1+\Omega_K)$ so that $\Omega_K \rightarrow 0$ implies $\eta\rightarrow \pm 1$, which means asymptotically flat FRW universe.}

 {Finally, \begin{equation}
    w_\phi=\frac{\text{sgn}(K)-\left(\text{sgn}(K)+3\right) \eta^2 -2 \lambda  y}{3 (\eta  x + y)}.  
\end{equation}}

 {We proceed with our analysis by considering $K=1$ and $K=-1$. \ Moreover, in the
following we assume that $\lambda=const$,  that is, we consider the exponential potential $V(\phi)=-{V_0} e^{\lambda  \phi }$, which leads to 
\begin{align}
    \phi = \frac{1}{\lambda}\ln \left(\frac{B}{6 \lambda   {V_0}}\right), \quad F(B) = \frac{B}{\lambda } \left(\ln
   \left(\frac{B}{6 \lambda   {V_0}}\right)
-1\right),  \label{pc.02}%
\end{align}
and to $h\equiv 0$.} 

\subsection{Positive curvature}

For an FLRW spacetime with positive spatial curvature, i.e. $K=1$, in the new
variables, the field equations become%
\begin{align}
\frac{dx}{d\tau}& =-1+4 \eta^2+\left(2 \eta^3+\eta\right) x+ y (2 \lambda +\lambda  \eta ) x-3),  \label{pc.01}%
\\
\frac{dy}{d\tau}& = y \left(\lambda  x+2 \eta \left(2 \eta^2+\lambda  y+1\right)\right), ~ \label{pc.02}%
\\
\frac{d\eta}{d\tau}& =\left(\eta ^2-1\right) \left(2 \eta^2+\lambda  y\right),  \label{pc.03}%
\end{align}
with constraint equation%
\begin{equation}
1+x\eta+y=0. \label{pc.04}%
\end{equation}

With the use of (\ref{pc.04}), we can write the equivalent
two-dimensional system%
\begin{align}
\frac{dx}{d\tau}&= -\left(\eta  x+1\right) \left(2 \lambda -2 \eta
   ^2+\lambda  \eta  x-4\right),  \label{pc.05}%
\\
\frac{d\eta}{d\tau}&=\left(1-\eta^2\right) \left(\lambda -2 \eta ^2+\lambda  \eta 
   x\right).  \label{pc.06}%
\end{align}
 {The observable quantities are reduced to 
\begin{equation}
 \{q, w_{tot},   w_\phi \}= \left\{2-\frac{\lambda(\eta  x+1)}{\eta
   ^2},1-\frac{2 \lambda  (\eta  x+1)}{3 \eta
   ^2},\frac{1}{3} \left(-2 \lambda +4 \eta ^2-2 \lambda 
   \eta  x-1\right)\right\}.
\end{equation}}

The equilibrium points $P=\left(  x\left(  P\right), \eta\left(  P\right)
\right)  $ of the dynamical system (\ref{pc.05}), (\ref{pc.06}) are given by
the algebraic equations%
\begin{align}
(\eta  x+1) \left(2 \lambda -2 \eta ^2+\lambda 
   \eta  x-4\right) &=0, \quad 
   \left(\eta ^2-1\right)
   \left(\lambda -2 \eta ^2+\lambda  \eta 
   x\right) =0,
\end{align}
that is%
\[
A_{1}=\left(  1, -1\right), ~A_{2}=\left(  -1, 1\right),
\]%
\[
~A_{3}=\left(  \frac{1}{\sqrt{\lambda}}, -\frac{\sqrt{\lambda}}{2}\right), ~A_{4}=\left(  -\frac{1}{\sqrt{\lambda}}, \frac{\sqrt{\lambda}}{2}\right),
\]%
\[
A_{5}=\left(  2-\frac{6}{\lambda}, -1\right), ~A_{6}=\left(  -2+\frac
{6}{\lambda}, 1\right).
\]

Points $A_{1}$,~$A_{2}$ describe exact solutions for which the kinetic part of the
scalar field dominates, i.e. $y\left(  A_{1}\right)  =y\left(  A_{2}\right)
=0$, with $w_{tot}\left(  A_{1}\right)  =w_{tot}\left(  A_{2}\right)  =1$ and
$q\left(  A_{1}\right)  =q\left(  A_{2}\right)  =2$. The spacetime is described
asymptotically by the spatially flat FLRW universe with scale factor
$a\left(  t\right)  =a_{0}t^{\frac{1}{3}}$.

Moreover, points $A_{3}$ and $A_{4}$ exist only when $\lambda>0$ and describe
Milne-like solutions with $a\left(  t\right)  =a_{0}t$. At these two points
the physical parameters are derived $y\left(  A_{3}\right)  =y\left(
A_{4}\right)  =-\frac{1}{2}$, $w_{tot}\left(  A_{3}\right)  =w_{tot}\left(
A_{4}\right)  =-\frac{1}{3}$ and $q\left(  A_{3}\right)  =q\left(
A_{4}\right)  =0$.

Finally the family of points $A_{5}$ and $A_{6}$ describe scaling solutions
with scale factor $a\left(  t\right)  =a_{0}t^{\frac{1}{\lambda-3}}$. We
derive $w_{tot}\left(  A_{5}\right)  =w_{tot}\left(  A_{6}\right)
=-3+\frac{2\lambda}{3}$ and $q\left(  A_{5}\right)  =q\left(  A_{6}\right)
=\lambda-4$, from which we observe that the exact solution describes an
accelerated universe when $\lambda<4$. The spatial curvature for the
background space is asymptotically zero. In the special case for which $\lambda
=3$, the exact solution at the equilibrium points is $a\left(  t\right)
=a_{0}e^{H_{0}t}$, which is the de Sitter solution.  {Evaluating at the equilibrium points $A_i$, we have $w_{tot}=w_{\phi}$.}

To investigate the stability properties of the equilibrium points we
determine the eigenvalues of the matrix%
\begin{equation}
\mathbf{A}=%
\begin{pmatrix}
\frac{\partial}{\partial x}\left(  \frac{dx}{d\tau}\right)  & \frac{\partial
}{\partial\eta}\left(  \frac{dx}{d\tau}\right) \\
\frac{\partial}{\partial x}\left(  \frac{d\eta}{d\tau}\right)  &
\frac{\partial}{\partial\eta}\left(  \frac{d\eta}{d\tau}\right)
\end{pmatrix}
_{\left(  x,y\right)  \rightarrow\left(  x\left(  P\right)  ,y\left(
P\right)  \right)  }.
\end{equation}
Let $e_{1}\left(  P\right)  $, $e_{2}\left(  P\right)  $ be the two
eigenvalues of the matrix $\mathbf{A}$.~We say that the equilibrium point
$P~$is an attractor and describes a stable asymptotic solution when the real
parts of the two eigenvalues are negatives. When the real parts of the eigenvalues are positive, point $P$ is called a source, and the asymptotic solution is an attractor. Otherwise, the equilibrium point $P$ is
characterized as a saddle point.

For each of the six equilibrium points, we derive the following set of
eigenvalues
\begin{align}
e_{1}\left(  A_{1}\right) &  = -4, ~e_{2}\left(  A_{1}\right)  =\lambda-6,
\\
e_{1}\left(  A_{2}\right) &  = 4, ~e_{2}\left(  A_{2}\right)  =-\left(
\lambda-6\right), \\
e_{1}\left(  A_{3}\right)   &  =\frac{1}{2} \sqrt{\lambda }
   \left(1+ \sqrt{(\lambda -8) \lambda +17} \right), 
~e_{2}\left(  A_{3}\right)    =\frac{1}{2} \sqrt{\lambda }
   \left(1-\sqrt{(\lambda -8) \lambda +17}\right),
\\
e_{1}\left(  A_{4}\right)   &  =-\frac{1}{2} \sqrt{\lambda }
   \left(1+ \sqrt{(\lambda -8) \lambda +17} \right), ~
e_{2}\left(  A_{4}\right)    =-\frac{1}{2} \sqrt{\lambda }
   \left(1-\sqrt{(\lambda -8) \lambda +17}\right),
\\
e_{1}\left(  A_{5}\right)  & =6-\lambda, ~e_{2}\left(  A_{5}\right)  =-2\left(
\lambda-4\right),
\\
e_{1}\left(  A_{6}\right) &  =-\left(  6-\lambda\right), ~e_{2}\left(
A_{5}\right)  = 2\left(  \lambda-4\right).
\end{align}

Points $A_{1,2}$ are nonhyperbolic when $\lambda=6$. 
Hence, point $A_{1}$ is a sink when $\lambda<6$; otherwise, it is a
saddle. Point $A_{2}$ is a source when $\lambda<6$; otherwise, it is a saddle
point. Points $A_{3}$ and $A_{4}\,$\ exist for $\lambda>0$, and they are nonhyperbolic for  $\lambda=4$, and a saddle otherwise. The equilibrium
points $A_{5,6}$ are nohyperbolic for $\lambda \in\{4,6\}$. We find that $A_5$ is  a sink for $\lambda >6$, a source when  $\lambda<4$ and a saddle for $4<\lambda <6$, while $A_6$ is  a source for $\lambda >6$, a sink when  $\lambda<4$ and a saddle for $4<\lambda <6$. 

\begin{figure}
    \centering
    \includegraphics[width=0.8\textwidth]{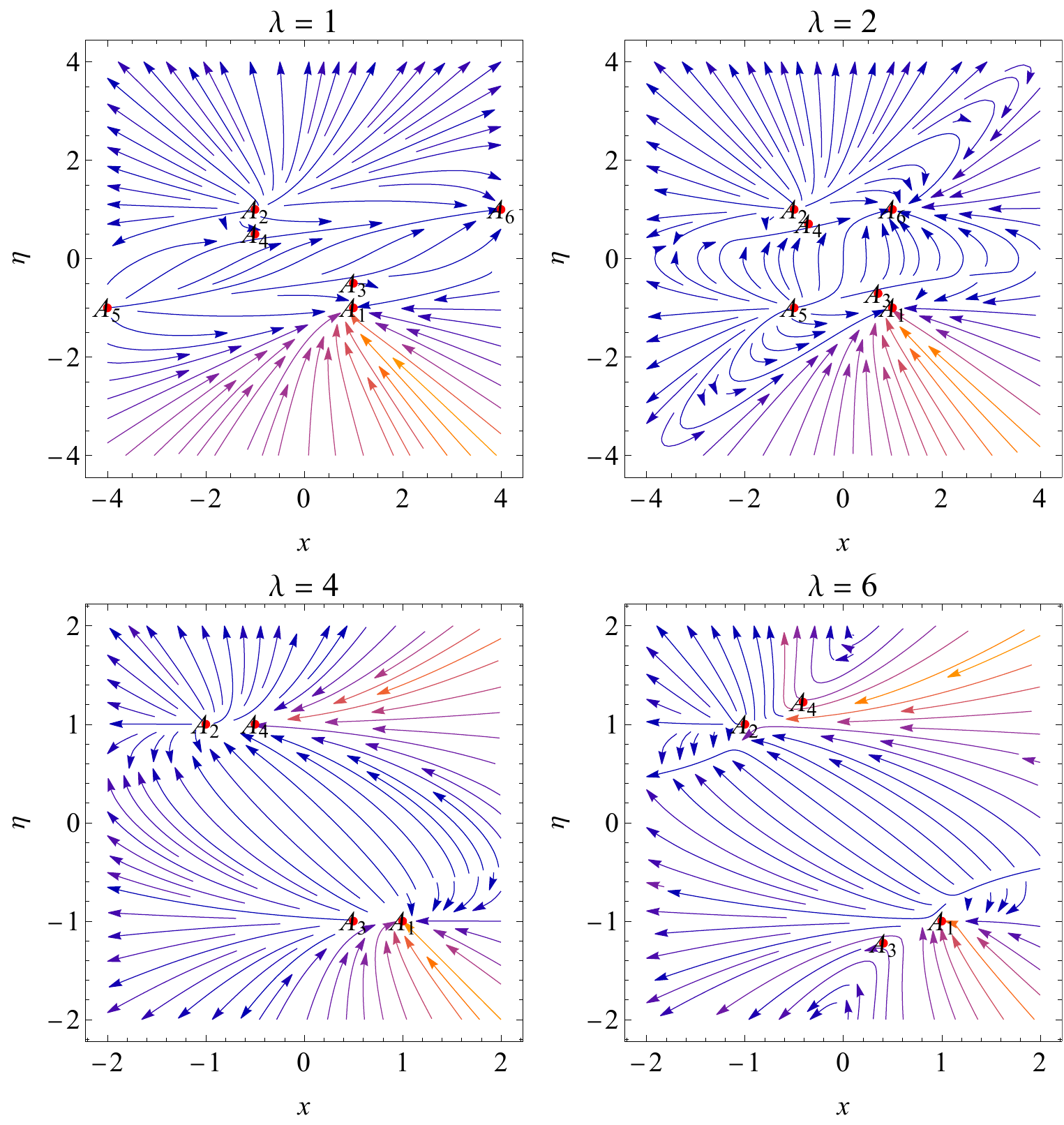}
    \caption{Phase plot of system \eqref{pc.05}, \eqref{pc.06} for $\lambda=1, 2, 4, 6$}
    \label{fig:A1}
\end{figure}

The results are summarized in Table \ref{tab1}. In figure \ref{fig:A1} is draw a phase plot of system \eqref{pc.05}, \eqref{pc.06} for $\lambda=1, 2, 4, 6$. 
\begin{table}[tbp] \centering
\caption{Asymptotic solutions for the field equations with positive spatial curvature.}%
\begin{tabular}
[c]{ccccc}\hline\hline
\textbf{Point} & \textbf{Curvature of FLRW} & $ {a}\left({t}\right)  $ & $w_\phi$ &
\textbf{Attractor?}\\\hline
$A_{1}$ & Flat & $t^{\frac{1}{3}}$ & $-1$ & $\lambda<6$\\
$A_{2}$ & Flat & $t^{\frac{1}{3}}$ & $-1$ & No\\
$A_{3}$ & $>0$ & $t$ & $-\frac{1}{3}$ & No\\
$A_{4}$ & $>0$ & $t$ & $-\frac{1}{3}$ & No \\
$A_{5}$ & Flat & $t^{\frac{1}{\lambda-3}},\lambda\neq3$& $-3+\frac{2\lambda}{3}$ & $\lambda>6$\\
&  & $e^{H_{0}t},\lambda=3$ & $-1$ & No \\
$A_{6}$ & Flat & $t^{\frac{1}{\lambda-3}},\lambda\neq3$ & $-3+\frac{2\lambda}{3}$ & $\lambda<4$\\
&  & $e^{H_{0}t},\lambda=3$ & $-1$ & Yes\\\hline\hline
\end{tabular}
\label{tab1}%
\end{table}%

\subsubsection{Poincaré variables}

To perform a complete analysis of the dynamics, we should investigate
if there exist equilibrium points when the dynamical variable, $x$, and $\eta$
take values at infinity.

Thus, we define the Poincaré variables%
\begin{equation}
x=\frac{X}{\sqrt{1-X^{2}-Z^{2}}}, ~\eta=\frac{Z}{\sqrt{1-X^{2}-Z^{2}}}
\label{pc.09}%
\end{equation}
and the new independent variable $d\sigma=\left(  1-X^{2}-Z^{2}\right)
^{-\frac{1}{2}}d\tau$.

\begin{table}[t] \centering
\caption{Asymptotic solutions at the infinity for the field equations with positive spatial curvature.}%
\begin{tabular}
[c]{cccc}\hline\hline
\textbf{Point} & $\left({X,Z}\right)  $ & ${a}\left({t}\right)  $ & \textbf{Stability}\\\hline
$P_{1}$ & $\left(  1,0\right)  $ & $a_{0}$  & Sink for $\lambda<0$, source for $\lambda>0$\\
$P_{2}$ & $\left(  -1,0\right)  $ & $a_{0}$ & Source for $\lambda<0$, sink for $\lambda>0$ \\\hline\hline
\end{tabular}
\label{tab2}%
\end{table}%

The two-dimensional dynamical system (\ref{pc.05}), (\ref{pc.06}) becomes%
\begin{align}
\frac{dX}{d\sigma}  &  = -2 \left(\lambda +(\lambda -2) X^4-2 (\lambda -1) X^3 Z+X^2
   \left(-2 \lambda +(2 \lambda -1) Z^2+4\right)-X Z \left(-2
   \lambda +(\lambda +2) Z^2+2\right)-(\lambda -1)
   Z^2-2\right),\label{eq.68}
\\
\frac{dZ}{d\sigma}  &  =\lambda -2 (\lambda -2) X^3 Z+X^2 \left(4 (\lambda
   -1) Z^2-\lambda \right)+X Z \left(3 \lambda +(2-4 \lambda )
   Z^2-4\right)+Z^2 \left(-3 \lambda +2 (\lambda +2)
   Z^2-2\right).\label{eq.69}
\end{align}

The equilibrium points of the latter system at infinity that is, on the
surface $1-X^{2}-Z^{2}=0$, are%
\[
P_{1}=\left(  1,0\right), ~P_{2}=\left(  -1,0\right).
\]%

Furthermore, in the new variables for the deceleration parameter, the
effective equation of state parameter, and the effective equation of state parameter of $\phi$,  we derive%
\begin{equation}
q\left(  X,Z\right)  =\lambda +\frac{\lambda  \left(X^2-X
   Z-1\right)}{Z^2}+2,
\end{equation}%
\begin{equation}
w_{tot}\left(  X,Z\right)  =\frac{2 \lambda  \left(X^2-X
   Z+Z^2-1\right)}{3 Z^2}+1,
\end{equation}
and 
\begin{equation}
w_\phi\left(  X,Z\right) =\frac{1}{3} \left(-2 \lambda +\frac{2
   \left(2 X^2+\lambda  X Z-2\right)}{X^2+Z^2-1}-5\right).
\end{equation}

Thus, points $P_{1}$ and $P_{2}$ describe static universes with$~a\left(
t\right)  =a_{0}$. We proceed with the study of the stability properties of the equilibrium points at infinity by using the parametrization
\begin{equation}
    X= (1-\varrho) \cos(\varphi), \quad Z= (1-\varrho)\sin(\varphi), \quad 0\leq \varrho <1,
\end{equation} and a time re-scaling
$d s=  d\sigma/(1-\varrho)$,
such  the region at infinite for $(x, \eta)$, i.e, $X^2 + Z^2=1$, is approached as $\varrho\rightarrow 0^ + $.

Taking the Taylor expansion centred in $\varrho=0$, neglecting higher order terms, we have
\begin{align}
\frac{d \varrho}{d s} &=\sin ^2(\varphi ) (\lambda  \cos (\varphi )-2 \sin (\varphi )), \quad 
\frac{d \varphi}{d s} =\frac{1}{2} \sin (\varphi ) (\lambda +2 (\lambda -3) \sin (2 \varphi )+(\lambda +2) \cos (2 \varphi )-2). \label{eq.76}
\end{align}
To find the equilibrium points at infinity, we solve the algebraic equation
\begin{align}
\sin ^2(\varphi ) (\lambda  \cos (\varphi )-2 \sin (\varphi ))=0, \quad 
\sin (\varphi ) (\lambda +2 (\lambda -3) \sin (2 \varphi )+(\lambda +2) \cos (2 \varphi )-2)=0. \label{eq.77}
\end{align}
Equations \eqref{eq.76} do not depend on the radial coordinate. Therefore,  the stability analysis considers the nature of the eigenvalues
\begin{equation}
    \Lambda_1(P)= 0, \quad
\Lambda_2(P)=\frac{1}{4} (-2 (\lambda -3) (\sin (\varphi )-3 \sin (3
   \varphi ))+(\lambda -6) \cos (\varphi )+3 (\lambda +2) \cos  (3 \varphi )). 
\end{equation}
of the Jacobian matrix evaluated at the values $\varphi$ that satisfy \eqref{eq.77}.

\begin{figure}
    \centering
    \includegraphics[width=0.8\textwidth]{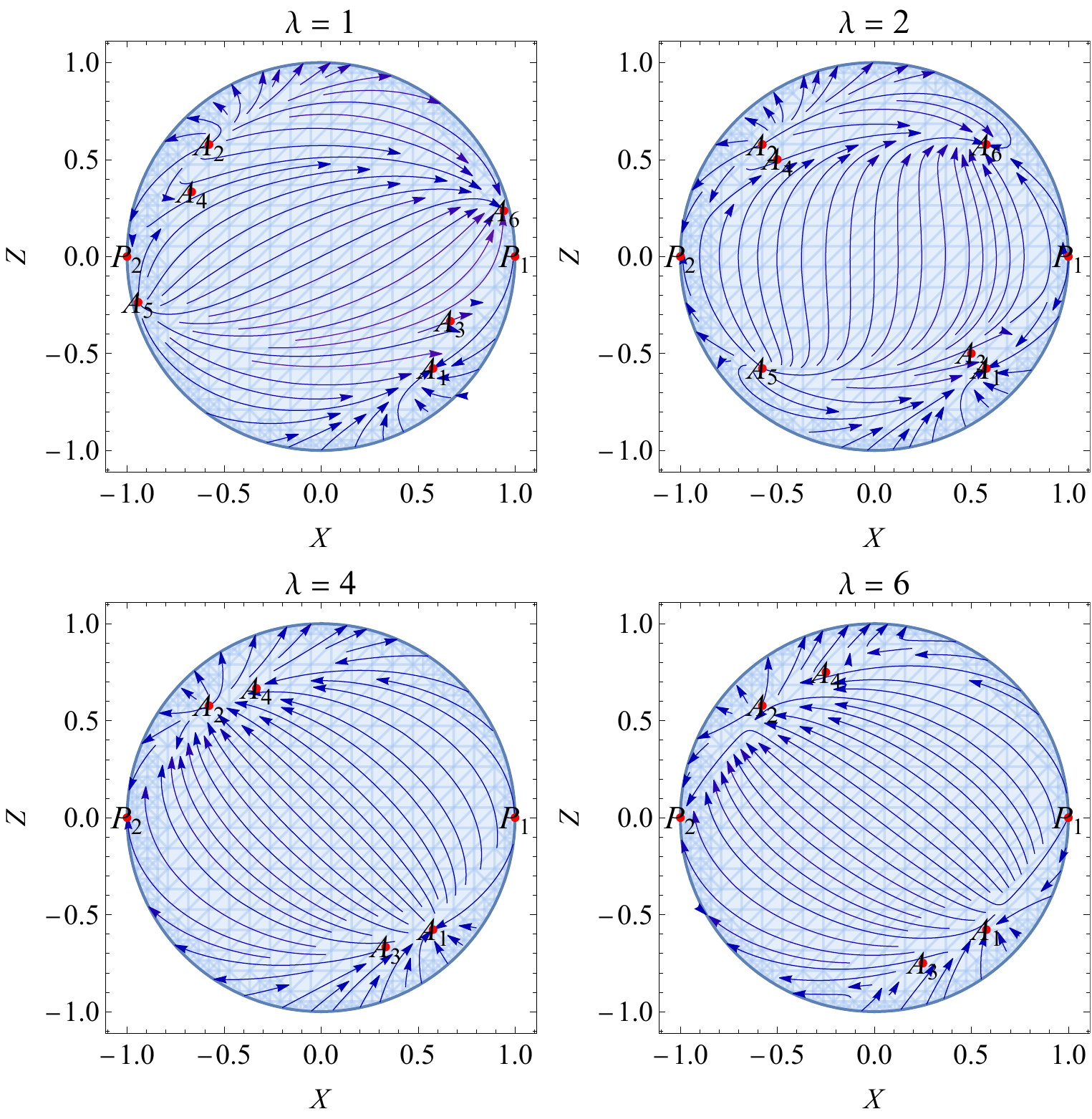}
    \caption{Phase plot of system \eqref{eq.68}-\eqref{eq.69} for $\lambda=1,2,4,6$}
    \label{fig:1}
\end{figure}

The solutions of \eqref{eq.77} are \newline
$P_1: =
 \varphi =  2 \pi  c_1\text{ if }c_1\in \mathbb{Z}$, with $\Lambda_1(P_1)=0, \Lambda_1(P_1)=\lambda$. Sink for $\lambda<0$, source for $\lambda>0$. \\
$P_2:=\varphi =  2 \pi  c_1+\pi \text{ if }c_1\in \mathbb{Z}$, with $\Lambda_1(P_2)=0, \Lambda_1(P_2)=-\lambda$. Source for $\lambda<0$, sink for $\lambda>0$. 
\newline
The results are summarized in Table \ref{tab2}.%

In figure \ref{fig:1} a phase plot of system \eqref{eq.68}-\eqref{eq.69} is presented for $\lambda=1,2,4$ and $6$. In the figures it is confirmed for $\lambda>0$ that $P_1$ is unstable and $P_2$ is stable.  Additionally, the information in table  \ref{tab1} relative to the stability of the points at the finite region $A_i, i=1 \ldots 6$ is confirmed.

\subsection{Negative curvature}

For an FLRW spacetime with negative spatial curvature, the field equations in
the dimensionless variables are%
\begin{align}
\frac{dx}{d\tau} & = \left(2 \eta^2+1\right) (\eta  x +1)+y  (2 \lambda  +\lambda  \eta x-3),  \label{pc.10}%
\\
\frac{dy}{d\tau} & = y  \left(\lambda  x +2 \eta   \left(2 \eta ^2+\lambda  y +1\right)\right),  \label{pc.11}%
\\
\frac{d\eta}{d\tau}& =\left(\eta ^2-1\right) \left(2 \eta ^2+\lambda  y\right), \label{pc.12}%
\end{align}
with constraint
\begin{equation}
\eta   (2 \eta  +x )+y -1=0. \label{pc.14}%
\end{equation}

Thus, with the use of the constraint equation (\ref{pc.14}), the dynamical system is reduced to the two-dimensional system%
\begin{align}
\frac{dx}{d\tau}&=\left(2 \eta  ^2+1\right) (\eta   x +1)-(\eta   (2
   \eta  +x )-1) (2 \lambda +\lambda  \eta    x -3),
\label{pc.15}%
\\
\frac{d\eta}{d\tau}&=\left(\eta  ^2-1\right) (\lambda -\eta   (2
   (\lambda -1) \eta  +\lambda  x )).  \label{pc.16}%
\end{align}
 {The observable quantities are reduced to 
\begin{align}
 \{q, w_{tot},   w_\phi \}   & =\left\{-2 \lambda +\frac{\lambda -\lambda  \eta   x }{\eta
    ^2}+2,\frac{1}{3} \left(-4 \lambda -\frac{2 \lambda  (\eta
     x -1)}{\eta  ^2}+3\right),\frac{2 \lambda +2 \eta  
   (\eta  -\lambda  (2 \eta  +x ))+1}{6 \eta
    ^2-3}\right\}.
\end{align}}

The equilibrium points for the dynamical system (\ref{pc.15}), (\ref{pc.16})
are
\[
B_{1}=\left(  1, -1\right), ~B_{2}=\left(  -1, 1\right),
\]%
\[
~B_{3}=\left(  \sqrt{\frac{2}{\lambda\left(  \lambda-2\right)  }}, -\sqrt
{\frac{\lambda}{2\left(  \lambda-2\right)  }}\right), ~B_{4}=\left(
-\sqrt{\frac{2}{\lambda\left(  \lambda-2\right)  }}, \sqrt{\frac{\lambda
}{2\left(  \lambda-2\right)  }}\right),
\]%
\[
B_{5}=\left(  2-\frac{6}{\lambda}, -1\right), ~B_{6}=\left(  -2+\frac
{6}{\lambda}, 1\right).
\]

The physical properties of the asymptotic solutions at the latter equilibrium
points are similar to those of points ${A}_i$. Indeed, points $B_{1}%
$,~$B_{2}$ describe spatially flat FLRW spacetimes with scale factor $a\left(
t\right)  =a_{0}t^{\frac{1}{3}}$, points $B_{3}$ and $B_{4}$ are real when
$\lambda\left(  \lambda-2\right)  >0$ and describe Milne universes, while for
the points $B_{5}$ and $B_{6}$ the asymptotic solution is that of spatially
flat FLRW with scale factor $a\left(  t\right)  =a_{0}t^{\frac{1}{\lambda-3}}%
$,~$\lambda\neq3~$or $a\left(  t\right)  =a_{0}e^{H_{0}t}$, $\lambda=3$.  {Evaluating at the equilibrium points $B_i$, we have $w_{tot}=w_{\phi}$.}

The eigenvalues of the linearized system (\ref{pc.15}), (\ref{pc.16}) around
the equilibrium points ${B}_i$ are calculated%
\begin{align}
e_{1}\left(  B_{1}\right) &  = -4, ~e_{2}\left(  B_{1}\right)  =\lambda-6,
\\
e_{1}\left(  B_{2}\right) & =4, ~e_{2}\left(  B_{2}\right)  =-\lambda+6,
\\
e_{1}\left(  B_{3}\right)   &  = \left(1-\sqrt{17-4 \lambda }\right)
   \sqrt{\frac{\lambda }{2(\lambda  -2)}},~ e_{2}\left(  B_{3}\right)  = \left(1+\sqrt{17-4 \lambda }\right)
   \sqrt{\frac{\lambda }{2(\lambda  -2)}},
\\
e_{1}\left(  B_{4}\right)   &  =-\left(1-\sqrt{17-4 \lambda }\right)
   \sqrt{\frac{\lambda }{2(\lambda  -2)}},
~e_{2}\left(  B_{4}\right)     =-\left(1+\sqrt{17-4 \lambda }\right)
   \sqrt{\frac{\lambda }{2(\lambda  -2)}},
\\
e_{1}\left(  B_{5}\right) & =6-\lambda, ~e_{2}\left(  B_{5}\right)  = -2\left(
\lambda-4\right)  ,
\\
e_{1}\left(  B_{6}\right) &  =-\left(  6-\lambda\right), ~e_{2}\left(
B_{6}\right)  =  2\left(  \lambda-4\right).
\end{align}

\begin{figure}
    \centering
    \includegraphics[width=0.8\textwidth]{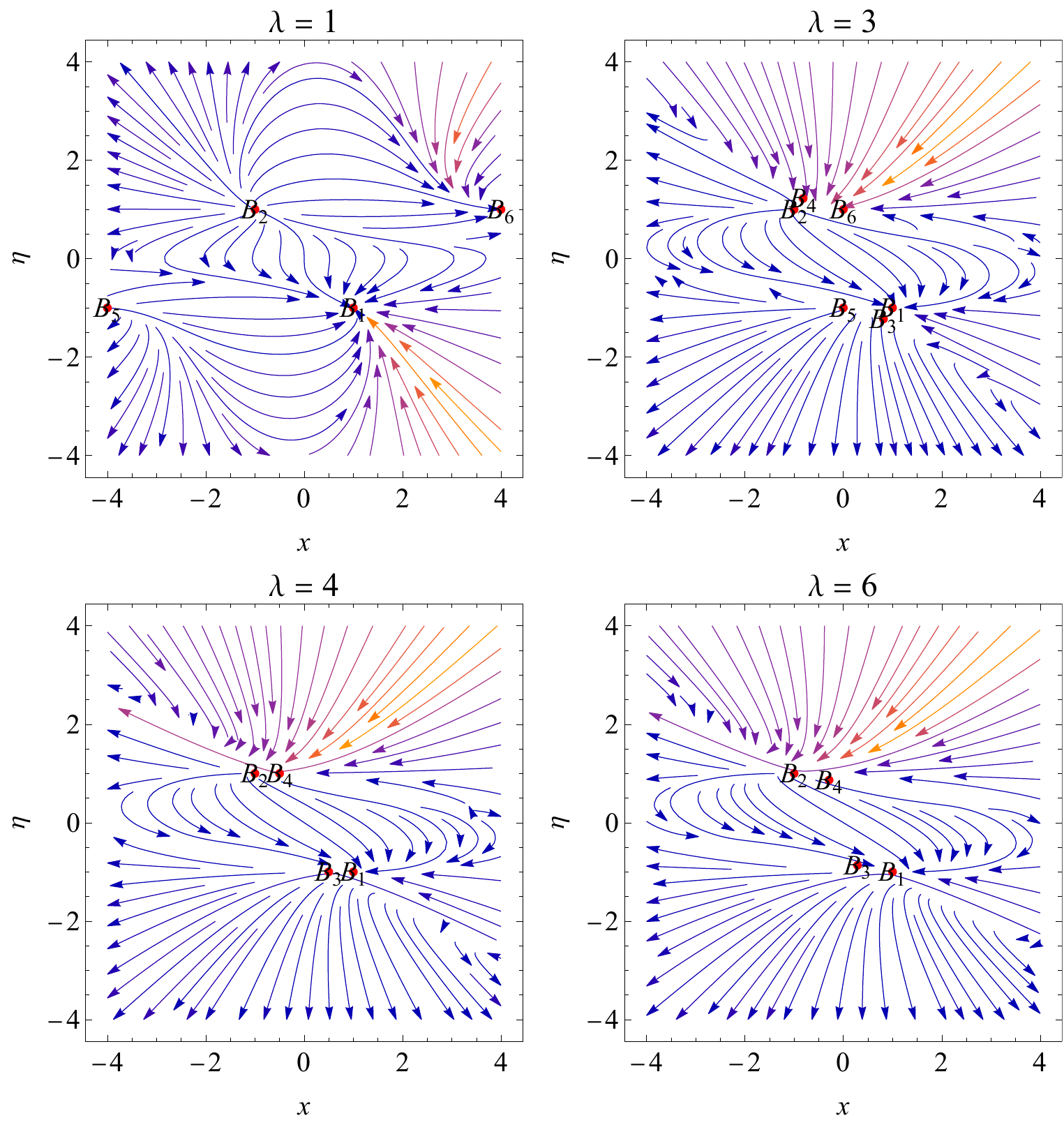}
    \caption{Phase plot of system \eqref{pc.05}, \eqref{pc.06} for $\lambda=1, 3, 4, 6$}
    \label{fig:B1}
\end{figure}

The stability properties of points $B_{1},~B_{2}$,~$B_{5}$ and $B_{6}$ are the
same as those of points $A_{1},~A_{2}$,~$A_{5}$ and $A_{6}$, respectively.
Point $B_{3}$ is a source for $4<\lambda \leq \frac{17}{4}$ (unstable node) or $\lambda>\frac{17}{4}$ (unstable spiral), or a saddle point for $\lambda <0$ or $2<\lambda <4$. On the other hand,
point $B_{4}$ is an attractor for $4<\lambda \leq \frac{17}{4}$ (stable node) or $\lambda>\frac{17}{4}$ (stable spiral), or a saddle point for $\lambda <0$ or $2<\lambda <4$. The results are summarized in Table \ref{tab3}. 
In figure \ref{fig:B1} is draw a phase plot of system  \eqref{pc.15}, \eqref{pc.16} for $\lambda=1, 3, 4, 6$. 
\begin{table}[tbp] \centering
\caption{Asymptotic solutions for the field equations with negative spatial curvature.}%
\begin{tabular}
[c]{ccccc}\hline\hline
\textbf{Point} & \textbf{FLRW} & ${a}\left({t}\right)  $ & $w_\phi$ &
\textbf{Attractor?}\\\hline
$B_{1}$ & Flat & $t^{\frac{1}{3}}$ & &$\lambda<6$\\
$B_{2}$ & Flat & $t^{\frac{1}{3}}$&  & No\\
$B_{3}$ & $<0$ & $t$&  & No \\
$B_{4}$ & $<0$ & $t$ & & $\lambda>4$\\
$B_{5}$ & Flat & $t^{\frac{1}{\lambda-3}},\lambda\neq3$  &  & $\lambda>6$\\
&  & $e^{H_{0}t},\lambda=3$ & & No\\
$B_{6}$ & Flat & $t^{\frac{1}{\lambda-3}},\lambda\neq3$ & & $\lambda<4$\\
&  & $e^{H_{0}t},\lambda=3$ & & Yes\\\hline\hline
\end{tabular}
\label{tab3}%
\end{table}%

\subsubsection{Poincaré variables}

We introduce the Poincaré variables (\ref{pc.09}), and we write the dynamical
system (\ref{pc.15}), (\ref{pc.16}) into the equivalent form%
\begin{align}
\frac{dX}{d\sigma}  &  = 2 \left(\lambda +(\lambda -1) X^4-2 X^3 Z+X^2 \left(-2
   \lambda +(2 \lambda -5) Z^2+2\right)+X Z \left((2-3 \lambda )
   Z^2+2\right)+(5-3 \lambda ) Z^2-1\right), \label{pc.20}%
\\
\frac{dZ}{d\sigma}  &  =\lambda  (2 Z (X-Z)+1)
   \left(X^2+X Z+3 Z^2-1\right)-2 Z \left(X^3+2 X^2 Z+X \left(5
   Z^2-1\right)-2 Z^3+Z\right). \label{pc.21}%
\end{align}
 {Furthermore, the physical variables in the new variables are
\begin{equation}
q\left(X,Z\right)  = -3 \lambda +\frac{\lambda -\lambda  X (X+Z)}{Z^2}+2,\label{pc.23}%
\end{equation}
\begin{equation}
w_{tot}\left(X,Z\right)  =\frac{2   \lambda -2 \lambda  X^2-2 \lambda  X Z+(3-6 \lambda ) Z^2}{3   Z^2},
\label{pc.22}%
\end{equation}
and
\begin{equation}
w_\phi\left(X,Z\right) = \frac{1}{9} \left(-6 \lambda +\frac{-4 X^2-6 \lambda  X   Z+4}{X^2+3 Z^2-1}+1\right).  
\end{equation}}

The equilibrium points for the dynamical system (\ref{pc.20}), (\ref{pc.21}) at
infinity, that is, on the surface $X^{2}+Z^{2}=1$, are%
\[
Q_{1}=\left(  1,0\right), ~Q_{2}=\left(  -1,0\right). 
\]%

Points $Q_{1},~Q_{2}$ describe static universe,$~a\left(  t\right)  =a_{0}$,
similarly with points $P_{1}$,~$P_{2}$.

To investigate the stability of the equilibrium points at infinite, we use the parameterization
\begin{equation}
    X= (1-\varrho) \cos(\varphi), \quad Z= (1-\varrho)\sin(\varphi),  \quad 0\leq \varrho <1,
\end{equation} and a time re-scaling
$d s=  d\sigma/(1-\varrho)$,
such as  the region at infinite for $(x, \eta)$, i.e, $X^2 + Z^2=1$, is approached as $\varrho$ tends to $0$.

Taking the Taylor expansion centred in $\varrho=0$ and neglecting error terms, we obtain that
\begin{align}
\frac{d \varrho}{d s} &=\sin ^2( \varphi ) (2 (\lambda -1) \sin ( \varphi)+\lambda  \cos ( \varphi )), \quad
\frac{d \varphi}{d s} =\frac{1}{2} \sin ( \varphi ) (5    \lambda +(4 \lambda -6) \sin (2  \varphi )+(8-3 \lambda ) \cos  (2  \varphi )-8). \label{eq.N76}
\end{align}
To find the equilibrium points at infinity, we solve the algebraic equation
\begin{align}
\sin ^2( \varphi ) (2 (\lambda -1) \sin ( \varphi)+\lambda  \cos ( \varphi ))=0, \quad 
\sin ( \varphi ) (5    \lambda +(4 \lambda -6) \sin (2  \varphi )+(8-3 \lambda ) \cos  (2  \varphi )-8)=0. \label{eq.N77}
\end{align}
In the first order, equations \eqref{eq.N76} do not depend upon the radial coordinate. Therefore, the stability analysis considers the nature of the eigenvalues
\begin{equation}
    \Lambda_1(Q)= 0, \quad 
\Lambda_2(Q)= \frac{1}{4} (-2 (2 \lambda -3) (\sin ( \varphi )-3 \sin
   (3  \varphi ))+(13 \lambda -24) \cos ( \varphi )+(24-9
   \lambda ) \cos (3  \varphi ))
\end{equation}
of the Jacobian matrix valuated at the values, $\varphi$, that satisfy \eqref{eq.N77}.
\begin{table}[t] \centering
\caption{Asymptotic solutions at the infinity for the field equations with negative spatial curvature.}%
\begin{tabular}
[c]{cccc}\hline\hline
\textbf{Point} & $\left({X,Z}\right)  $ & ${a}\left({t}\right)  $ & \textbf{Stability}\\\hline
$Q_{1}$ & $\left(  1,0\right)  $ & $a_{0}$ & Sink for $\lambda<0$, source for $\lambda>0$\\
$Q_{2}$ & $\left(  -1,0\right)  $ & $a_{0}$ & Source for $\lambda<0$, sink for $\lambda>0$\\
\hline\hline
\end{tabular}
\label{tab4}%
\end{table}%

\begin{figure}
  \includegraphics[width=0.8\textwidth]{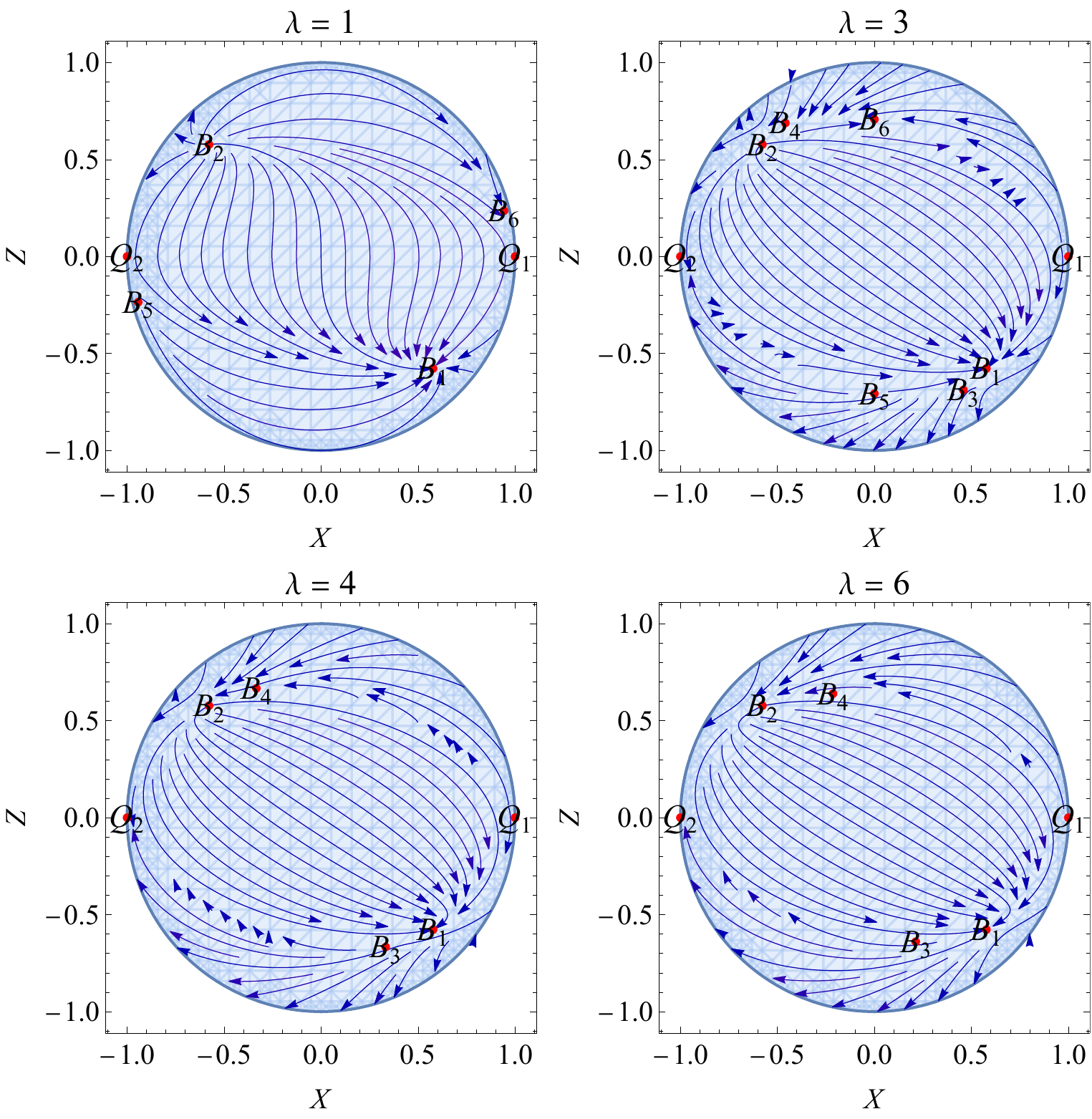}
    \caption{Phase plot of system \eqref{pc.20}-\eqref{pc.21} for $\lambda=1, 3, 4, 6$}
    \label{fig:2}
\end{figure}
The solutions of \eqref{eq.N77} are  \newline
$Q_1: =
 \varphi =  2 \pi  c_1\text{ if }c_1\in \mathbb{Z}$, with $\Lambda_1(Q_1)=0, \Lambda_1(Q_1)=\lambda$. Sink for $\lambda<0$, source for $\lambda>0$. \\
$Q_2:=\varphi =  2 \pi  c_1+\pi \text{ if }c_1\in \mathbb{Z}$, with $\Lambda_1(Q_2)=0, \Lambda_2(Q_2)=-\lambda$.  Source for $\lambda<0$, sink for $\lambda>0$.\\
\newline
The dynamics at infinity ($\varrho=0$) are governed by the one-dimensional dynamical system \eqref{eq.N76}.
In figure  \ref{fig:2} a phase plot of system \eqref{pc.20}-\eqref{pc.21} is presented for $\lambda=1, 3, 4, 6$. In the figures, it is confirmed that for $\lambda>0$, $Q_1$ is unstable and $Q_2$ is stable. Moreover, the information in table  \ref{tab3} relative to the stability of the points at the finite region $B_i, i=1 \ldots 6$ is confirmed. The stability of the rest of the points at the infinite region is as discussed in table \ref{tab4}.

\section{Conclusions}

\label{sec5}

In this piece of work, we studied the cosmological model of the fourth-order
teleparallel theory gravity known as $f\left(  T,B\right)  $ in the case of
FLRW background space with nonzero spatial curvature. In particular, we assumed
that $f\left(  T,B\right)  $ is a linear function of\ $T$, $f\left(
T,B\right)  =T+F\left(  B\right)  $, such that the modifications of the
gravitational Action Integral of the TEGR to be introduced by the term
$F\left(  B\right)  $. In the case where the function $F\left(  B\right)  $ is linear, the TEGR is recovered; thus, in this work, we consider $F\left(
B\right)  $ to be a nonlinear function.

For the proper vierbein fields, we derived the field equations for the
background space of our consideration. Using Lagrange multipliers,
we introduced a scalar field that attributes the
geometrodynamical degrees of freedom to writing the field equations
in the form of second-order theory. The dynamical variables are the scale factor
$a\left(  t\right)  $ and the scalar field $\phi\left(  t\right)  $. Moreover,
we show that this cosmological model admits a minisuperspace description.
Thus, there exists a point-like Lagrangian, the variation of which provides the field
equations. That specific characteristic of the gravitational theory is
essential because various techniques from analytic mechanics can be applied
for the investigation of the differential equations, also, the Wheeler-DeWitt
equation of quantum cosmology can be calculated straightforward.

We investigated the existence of exact solutions where the scale factor is a
power-law function, scaling solution, or exponential function. These 
exact solutions are essential because they can describe specific eras of
cosmological history. Moreover, we proved that both cases have a
scalar field potential, so the field equations can be solved explicitly.

Assume now, an arbitrary scale factor $a\left(  t\right)  $, then from
equations (\ref{fc.18}) and (\ref{fc.20}) it follows
\begin{equation}
\ddot{\phi}-H\dot{\phi}-2Ka^{-2}+2\dot{H}=0,
\end{equation}
which is a second-order equation of the form%
\begin{equation}
\ddot{\phi}+\alpha\left(  t\right)  \dot{\phi}+\beta\left(  t\right)  =0
\label{pc.30}%
\end{equation}
with $\alpha\left(  t\right)  =-H$ and $\beta\left(  t\right)  =-2Ka^{-2}%
+2\dot{H}$. Equation (\ref{pc.30}) is a linear equation and it is maximally
symmetric, which means that it always admits a solution for an arbitrary function
$\alpha\left(  t\right)  $ and $\beta\left(  t\right)  $, that is, for
arbitrary selection of the scale factor $a\left(  t\right)  $. Hence, it is
easy to infer that the field equations in these cosmological scenarios are
always integrable. Such analysis generalizes previous results on this theory in
the case of a spatially flat FLRW geometry.

Finally, we studied the general evolution of the dynamical variables 
described by the field equations. In particular, we derived the equilibrium
points and investigated their stability properties. We performed our analysis
separately for the cases of positive and negative spatial curvature. The
equilibrium points were studied in the finite and infinite regions using Poincaré variables. While, from a first read, it seems that there are
similarities in the cosmological evolution for the two cases of positive and
negative spatial curvature, from the detailed analysis, we found that the
results differ.

The results of this work are essential to understanding the spatial
curvature in teleparallelism. The $f\left(  T, B\right)  $ theory
provides important asymptotic behaviours of particular interest. In future
work, we plan to study further the applications of $f\left(  T, B\right)  $
theory in the pre-inflationary era.

\section{Conflict of Interest Statement} 

The authors declare to have no conflict of interest.

\section{Author Contributions}

The authors equally contributed.

\section{Acknowledgements}
This work is based on the research supported in part by the National Research Foundation of South Africa (Grant Numbers 131604). Additionally, this research is funded by Vicerrector\'ia de Investigaci\'on y Desarrollo Tecnol\'ogico (Vridt) at Universidad Cat\'olica del Norte.  G. L. was funded through Concurso De Pasant\'ias De Investigaci\'on Año 2022, Resolución Vridt No. 040/2022 and through Resolución Vridt No. 054/2022.


\end{document}